\documentclass{hep99}
\usepackage{epsf}
\def\tb{\bar t}
\def\bb{\bar b}

\def\to{\rightarrow}

\def\be{\begin{equation}}
\def\ee{\end{equation}}
\def\bea{\begin{eqnarray}}
\def\eea{\end{eqnarray}}

\begin{document}
\renewcommand{\topfraction}{0.99}
\renewcommand{\bottomfraction}{0.99}
\topmargin=-0.3in
%\twocolumn[\hsize\textwidth\columnwidth\hsize\csname @twocolumnfalse\endcsname

\title{Probing Anomalous Top Quark Couplings at the Future Linear
Colliders$^*$
}

\author{T. Huang$^\dag$ \ftnote{1}\\}

\address{CCAST (world laboratory), P.O. Box 8730, Beijing 100080,
China\\
Institute of High Energy Physics, Chinese Academy of Sciences,\\ P.O. Box 918, Beijing, 100039, China}

\abstract{
In terms of an effective Lagrangian we investigate
the possibilities of probing anomalous top
quark couplings, $t \bar{t} H$, $\gamma t \bar{t}$, $Z t \bar{t}$
 and $tWb$ at the future linear colliders.
% We examine firstly the constraints on these anomalous couplings
%from the $Z\to b \bar{b}$ data at LEP I
 %and unitarity considerations. Secondly we study in detail the processes
%$e^+e^- \to t \bar{t} H$.
It is found that
at a linear collider with a c.~m.~energy
$\sqrt s \sim 0.5-1.5$ TeV and a high luminosity of $10-1000$
fb$^{-1}$, $e^+e^- \to t \bar{t} H$ is an ideal process in probing
anomalous $t \bar t H$ couplings.
%Thirdly we investigate process $e^+ e^-
%\to t \bar{t}$.
We also study in detail
the effects of anomalous couplings on $t \bar{t}$ spin correlations in the
top pair production as well as the top quark decay
 processes with three bases (helicity, beam line and off-diagonal bases).
Our results show that
with a c.~m.~energy
 $\sqrt s \sim 0.5-1$ TeV and a high luminosity of $1-100$  fb$^{-1}$,
the anomalous couplings  $\gamma t \bar{t}$, $Z t\bar{t}$
 and $tWb$ may be sensitively probed.
}
\maketitle
%]
\noindent{
\fntext{}{--------------------------------}
\fntext{1}{$^*$Invited talk at the XXX International Conference on High Energy Physics, Osaka, July 26-August 2,
2000.}

\fntext{2}{$^\dag$Written with T. Han, Z.-H. Lin, J.-X. Wang and X.
Zhang.}      }
% \normalsize
% \vfill
% \vspace*{.5cm}
% \bigskip
% \vfill
% \footnoterule
% \noindent
% {\small
%$^\ast$Invited talk at the XXX International Conference on High Energy Physics, Osaka, July 26-August 2, 2000.}
%\\
%$^{\dag}$~{\small Written with T. Han, Z.-H. Lin, J.-X. Wang and X. Zhang. }
%\bigskip     ]
%%%%%%%%%%%%%%%%%%%%%%%%%%%%%%%%%%%%%
%\section{Introduction}
One believes that the large top-quark mass, which is close to the
order of the weak scale ($m_t\approx v/\sqrt{2}$), makes the third generation to play a
significant role in probing the new physics beyond the
Standard Model (SM). Thus the linear collider (LC) will have a potential to explore the new
physics associated with the Higgs and the top-quark sector.

In order to explore the possibility, we take a model-independent approach by
using a linearly realized effective Lagrangian to dimension-6 operators
including CP violation. We discuss the process $e^+e^- \to t \bar{t} H$
and the top quark spin correlation to probe the non-standard couplings.
Particularly, the process $e^+e^- \to t \bar{t} H$ is an ideal one
for probing anomalous coupling $t\tb H$ and hopefully gains some insight for the new physics beyond
the SM. The observability of the signal from anomalous couplings
$t \bar{t} H$, $\gamma t \bar{t}$, $Z t \bar{t}$ and $tWb$ with
 c.~m.~energy $\sqrt s \sim 0.5-1$ TeV is studied.
%%%%%%%%%%%%%%%%%%%%%%%%%%%%%%%%%%%%%%%
%\section{Effective interactions}
\vspace*{.5cm}

 In the case of linear realization, the new physics is parameterized by
 higher dimensional operators which contain the SM fields and are
 invariant under the SM gauge group.
 Below the new physics scale $\Lambda$, the effective Lagrangian
 can be written as
 \begin{equation}
 \label{eff}
 {\cal L}_{eff}={\cal L}_0+\frac{1}{\Lambda^2}\sum_i C_i O_i
                          +{\cal O}(\frac{1}{\Lambda^4})
                          \end{equation}
  where ${\cal L}_0$ is the SM Lagrangian.
  $O_i$ are dimension-6 operators which are
  $SU_c(3)\times SU_L(2)\times U_Y(1)$ invariant and $C_i$ are coefficients
which represent the coupling strengths of $O_i$ \cite{linear}.
%Recently the effective operators involving the top quark were
%reclassified and some are analyzed in Refs.~\cite{Whisnant,Renard}.
All the operators $O_i$ are hermitian and the coefficients $C_i$
are real and the order of unity.
If we assume that the new physics is of the origin associated with
the electroweak symmetry breaking, then it is natural to identify
the cut-off scale $\Lambda$ to be the order of ${\cal O}(4\pi v)$. Alternatively,
based on unitarity argument for massive quark scattering \cite{topu},
the scale for new physics in the top-quark sector
should be below about 3 TeV.
There are twelve dimension-six CP even operators. All the operators, which
give new contributions to the couplings of $t \bar{t} H$, $\gamma t \bar{t}$, $Z t \bar{t}$ and $tWb$
, are listed in Refs.~\cite{Whisnant,tth,spin}.

Among them, some of the operators are energy independence such as
\bea
O_{t1}=(\Phi^{\dagger}\Phi-\frac{v^2}{2})\left [\bar q_L
         t_R\widetilde\Phi
         +\widetilde\Phi^{\dagger} \bar t_R q_L\right ]
\eea
and some are energy-dependent, such as
\bea
O_{Dt}=(\bar q_L D_{\mu} t_R) D^{\mu}\widetilde\Phi
         +(D^{\mu}\widetilde\Phi)^{\dagger}(\overline{D_{\mu}t_R}q_L)
\eea
due to the deviative. The energy-dependence of all dimension-6 operators are listed
in the table of Refs.~\cite{Whisnant,tth,spin}.
%%%%%%%%%%%%%%%%%%%%%%%%%%%%%%%%%%%%%%
%\section{Bounds on the nonstandard couplings}
\vspace*{.5cm}

Generally, we can examine the possible constraints on
the operators from the measurement $Z\to b \bb$. The observable
$R_b$ at the $Z$ pole is calculated to be
\begin{eqnarray}\label{Rb}
R_b &\equiv& {\Gamma(Z\to b\bar b)\over \Gamma(Z\to {\rm hadrons})} \nonumber \\
&=&  R_b^{SM}\left[ 1+2\frac{v_b\delta V+a_b\delta A}{v_b^2+a_b^2}
      (1-R_b^{SM})\right ],
\end{eqnarray}
where $v_b$ and $a_b$ represent the SM couplings
and $\delta V,\ \delta A$ the new physics contributions.
If we attribute the difference between $R_b^{SM}$ and $R_b^{exp}$ as the new physics contribution,
we obtain the limit at the $1\sigma$
($3\sigma$) level
\begin{equation}\label{bound1}
-4\times 10^{-3}~ (-8\times 10^{-3})<\delta V
<-5\times 10^{-5}~ (4\times 10^{-3}).   \nonumber \\
\end{equation}
Assuming that there is no accidental
cancellation between different operators and
noting that\\ $2s_Wc_W m_Z/ev \simeq 1$,
we obtain the bound for each of them at the $1\sigma$
($3\sigma$) level as
\begin{eqnarray}\label{bound2}
5\times 10^{-5}~ (-4\times 10^{-3})&< &
\frac{v^2}{\Lambda^2}C^{(1)}_{\Phi q}
~(or~\frac{v^2}{\Lambda^2}C^{(3)}_{\Phi q}) \nonumber \\
&<&4\times 10^{-3}~ (8\times 10^{-3}).
\end{eqnarray}
On other hand, the constraints from
$A_{FB}^{(b)}$ are weaker than $R_b$.

For $O_{t1}$, $O_{t2}$, $O_{Dt}$, $O_{tW\Phi}$ and $O_{tB\Phi}$
they are not constrained by $R_b$ at tree level and
bounds on them can be studied
from the argument of partial wave unitarity in Ref.~\cite{Renard}.
It is informative to see the ranges
of the unitarity bounds for $\Lambda \approx 3-1$ TeV:
\begin{eqnarray}
&& |C_{t1}|{v^2\over \Lambda^2} \simeq 1.0 - 3.0,\
\label{ft1} \qquad  \qquad
|C_{t2}|{v^2\over \Lambda^2} \simeq 0.29 - 2.6, \nonumber \\
&& C_{Dt}{v^2\over \Lambda^2} \simeq 0.07-0.63  \ \ \mbox{or } \ \
C_{Dt}{v^2\over \Lambda^2} \simeq -(0.04-0.40), \nonumber \\[0.1cm]
&& |C_{tW\Phi}|{v^2\over \Lambda^2} ~or~
|C_{tB\Phi}|{v^2\over \Lambda^2} \simeq 0.02 - 0.15\ .
\end{eqnarray}

At present, there is no significant experimental constraint on
the CP-odd couplings involving the top-quark sector.
%%%%%%%%%%%%%%%%%%%%%%%%%%%%%%%%%%%%%%%%%%%%%%
%\section{$t\bar t H$ Production with anomalous Couplings}
\vspace*{.5cm}

The relevant Feynman diagrams for $e^+e^- \to t\bar t H$
production are depicted in Fig.~1, where
(a)$-$(c) are those in the SM and the dots denote the contribution
from new interactions. The four-particle vertex (Fig.~1d)
should be paid more attention since there is no such vertex in the SM
but exists in the effective couplings due to the gauge invariance.
We evaluate all the diagrams
including interference effects, employing a helicity
amplitude package (FDC) developed in \cite{Wang}. This package
has the flexibility to include new interactions beyond
the SM. We have not included the QCD corrections to the
signal process, which are known to be positive and
sizable.
\begin{figure}[thb]\label{feyn}
\centerline{\epsfysize 0.7 truein \epsfbox{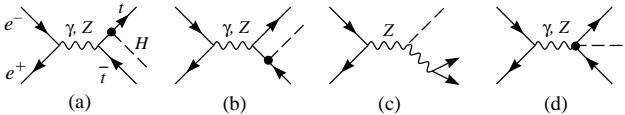}}
\caption[]{Feynman diagrams for $e^+e^- \to t\bar t H$ production.
(a)-(c) are those in the SM. The dots denote the contribution
from new interactions.}
\end{figure}

Due to the strong constraints on $O_{\Phi q}^{(1)}$ and
$O_{\Phi q}^{(3)}$ from the $Z\to b\bar b$ measurement,
(see Eq.~(\ref{bound2})),
the effects of these operators at colliders will be rather small
and can be neglected.
Following the energy-dependence behavior, we expect that modifications to
the SM prediction from different operators would be distinctive at high energies.
For the purpose of illustration, we only present results for the
operators $O_{t1}$ (energy-independent) and $O_{Dt}$ (most sensitive to
energy scale) to represent to others.

The results include the production cross sections versus $\sqrt{s}$,
the Higgs mass $m_H$ and the couplings. The effect due to the operator
$O_{Dt}$ is insignificant at $\sqrt{s}=0.5$ TeV, which at higher energies
the contribution from $O_{Dt}$ is substantial.

To establish the sensitivity limits on the anomalous couplings that
may be probed at the future LC experiments,
one needs to consider the identification of the final state from $t\tb H$,
including the branch ratios and the detection efficiencies.
The branching ratio of the leading decay mode $H\to b \bb$ is about
$80\% \sim 50\%$ for the mass range of $100\sim 130$ GeV. We assume
$65\%$ efficiency for a single b-tagging to identify four b-jets in the final
state. With the desirable consideration we estimate an efficiency factor
$\epsilon_S$ for detecting $e^+e^- \to t \bar{t} H$ to be
$\epsilon_S=10-30\%$ and a factor $\epsilon_B$ for reducing QCD and EW
background to be $\epsilon_B=10\%$. In order to estimate the luminosity (L)
needed for probing the effects of the non-standard couplings,
we define the significance of a signal rate (S) relative to a background
rate (B) in terms of the Gaussian statistics
\bea
\sigma_S=\frac{S}{\sqrt{B}}
\eea
for which a signal at 95\% (99\%) confidence level (C.L.) corresponds to $\sigma_S=2~(3)$.
They are calculated as
\bea\label{sb}
S&=&L(|\sigma-\sigma_{SM}|)\epsilon_S,\nonumber \\
B&=&L\left [\sigma_{SM}
\epsilon_S+(\sigma_{QCD}+\sigma_{EW})\epsilon_B\right ].
\eea
Then we obtain the luminosity required for observing the effects of $O_{t1}$ and $O_{Dt}$ at 95\%
C.L. for 0.5 TeV and 1 TeV for $C_{t1}$ and for 1 TeV and 1.5 TeV for $C_{Dt}$ in Fig.~2
where the two curves are for 10\% and 30\% of signal detection efficiency, respectively.
It can be seen that at a 0.5 TeV collider, one would need rather high integrated luminosity
to reach the sensitivity to the anomalous couplings; while at a collider with a higher c.m. energy
one can sensitively probe those couplings with a few hundred $fb^{-1}$ luminosity.
\begin{figure}[thb]
\centerline{\epsfysize 2 truein \epsfbox{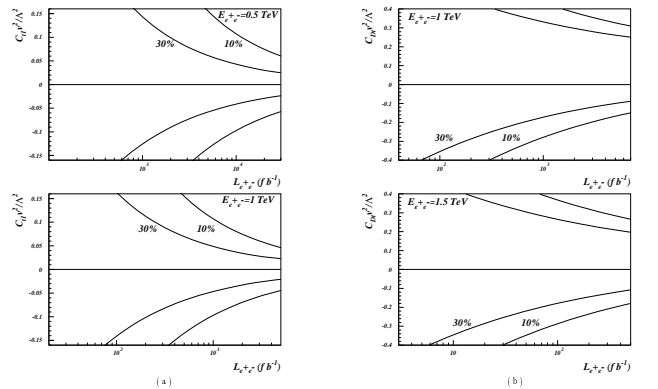}}
\caption[]{(a) Sensitivity to the anomalous couplings
versus the integrated luminosity for a 95\% confidence level limits
with $m_H=120$ GeV for (a) $O_{t1}$  at $\sqrt s=0.5$ TeV and $\sqrt s=1$
TeV and for (b) $O_{Dt}$ at $\sqrt s=1$ TeV and $\sqrt s=1.5$ TeV.
\label{lumi}}
\end{figure}

%\begin{figure}[thb]
%\centerline{\epsfysize 4.5 truein \epsfbox{fig5.eps}}
%\caption[]{Sensitivity to the anomalous couplings $O_{t1}$
%versus the integrated luminosity for a 95\% confidence level limits
%at (a) $\sqrt s=0.5$ TeV and (b) $\sqrt s=1$ TeV, with $m_H=120~GeV$.
%\label{lumi}}
%\end{figure}
%
%\begin{figure}[thb]
%\centerline{\epsfysize 4.5 truein \epsfbox{fig6.eps}}
%\caption[]{Sensitivity to the anomalous couplings $O_{Dt}$
%versus the integrated luminosity for a 95\% confidence level limits
%at (a) $\sqrt s=1$ TeV and (b) $\sqrt s=1.5$ TeV, with $m_H=120~GeV$.
%\label{lumi2}}
%\end{figure}
%

If there exist effective CP-odd operators beside the SM interaction,
then CP will be violated in the Higgs and top-quark sector. By using
the similar discussion, one can try to observe the effects of the operators
beyond the SM expectation. The CP-violating effect can be parameterized
by a cross section asymmetry as
\begin{eqnarray}
A_{CP}\equiv \frac{\sigma((p_1\times p_3)\cdot p_4<0)-
\sigma((p_1\times p_3)\cdot p_4>0)}
{\sigma((p_1\times p_3)\cdot p_4<0)+
\sigma((p_1\times p_3)\cdot p_4>0)}
\end{eqnarray}
where $p_1$, $p_3$ and $p_4$ are the momenta of the incoming
electron, top quark and anti-top quark, respectively.
The luminosity required for detecting the effects on the total
cross sections and $A_{CP}$ is shown in Fig.~3 versus CP-odd
couplings with 95\% C.L. for $m_H=120$ GeV and $\sqrt{s}=1$ TeV.
The solid curves are for the cross sections with efficiency
factor $\epsilon_S=30\%$ and $\epsilon_B=10\%$ according to Eq.~(\ref{sb}).
Apparently, the effects on the total cross section due to CP-odd operators
are much stronger than that on $A_{CP}$. In other words, the
direct observation of the CP asymmetry would need much lighter luminosity to reach.
\begin{figure}[thb]
\centerline{\epsfysize 3 truein \epsfbox{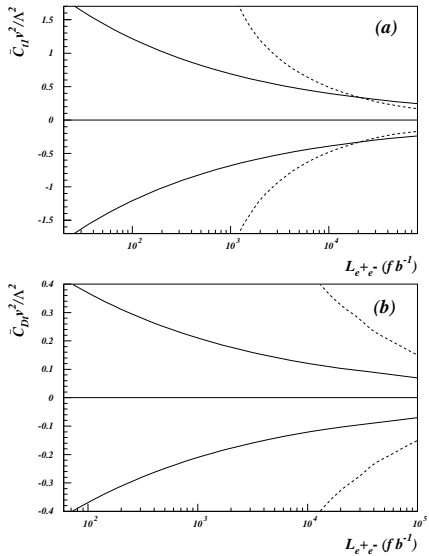}}
\caption[]{
Sensitivity to the anomalous CP-odd couplings
versus the integrated luminosity for a 95\% confidence level limits
and for $30\%$ of detection efficiency
at $\sqrt s=1$ TeV , with $m_H=120$ GeV.
The solid line is for the total cross section and
the dash line is for the CP asymmetry $A_{CP}$.
\label{cp3}}
\end{figure}
%
%%%%%%%%%%%%%%%%%%%%%%%%%%%%%%%%%%%%%%%%%%%%%%%%%
%\section{Top quark spin correlation}
\vspace*{.5cm}

We now discuss the corrections from the anomalous couplings in the top-quark
pair production and the top-quark decays. In order to evaluate the possible effects
from new physics and study which spin basis is more sensitive to anomalous couplings,
we use the generic spin basis suggested by Parke and Shadmi~\cite{parke}.
They are helicity basis, beamline basis and off-diagonal basis. With the above bases
and center of mass energy $\sqrt{s}$, we calculate the differential polarized cross section
at the tree level and the differential cross section of
$e^-e^+ \to t\tb \to (b \bar l \nu_{\bar l})(\bar b l' \nu_{l'})$.
In order to see the effects of the anomalous coupling, we
define $<S_t>$, $<S_{\tb}>$ and $<S_t S_{\tb}>$ are $t$, $\tb$ and
$t\tb$ correlation functions and calculate the relevant observables.
Our results show that with a c.m. energy $\sqrt{s}\sim 0.5-1$ TeV and a higher
luminosity of $1-100$ $fb^{-1}$, the anomalous couplings $\gamma t \tb$,
$Z t \tb$ and $tWb$ may be sensitively probed~\cite{spin}.
%%%%%%%%%%%%%%%%%%%%%%%%%%%%%%%%%%%%%%%%%%%%%%%%
%\section{Conclusions}
\vspace*{.5cm}

In summary, we have considered a general effective lagrangian to dimension-6 operators
including CP violation effects. The constraints on these anomalous couplings
has been derived from the $Z\to b \bb$ data and unitarity consideration.

In order to explore the effects of these non-standard couplings,
we have studied the process
$e^+e^- \to t \bar{t} H$, anomalous couplings
($\gamma t \bar{t}$, $Z t \bar{t}$ and $tWb$) on $t\tb$ spin correlations
in the top pair production as well as the top-quark decay process with three bases
(helicity, beamline and off-diagonal basis). We find that the future linear collider
experiments should be able to probe those couplings well below their unitarity bounds.
To reach a good sensitivity, the higher integrated luminosity needed is about
several hundred
$fb^{-1}$ for a c.m. $\sqrt{s}\sim 0.5-1.5$ TeV.
%%%%%%%%%%%%%%%%%%%%%%%%%%%%%%%%%%%%%%%%%%%%%%%%

%%%%%%%%%%%%%%%%%%%%%%%%%%%%%%%%%%%%%%%%%%%%%%%%%%
\end{document}